\documentclass[aps,prl,reprint]{revtex4-1}
\usepackage{verbatim}
\usepackage{simplewick}
\usepackage{graphicx,color} 
\usepackage{booktabs} 
\usepackage{color}
\usepackage[normalem]{ulem}
\usepackage[matrix,frame,arrow]{xy}

\graphicspath{{./figures/}}
\usepackage{hyperref}
\newcommand{\titlename}{Unified views of quantum simulation algorithms for chemistry }

\hypersetup{
  pdftitle={\titlename},
  colorlinks,%
  citecolor=black,%
  filecolor=black,%
  linkcolor=black,%
  urlcolor=black,
  pdfauthor={J. D. Whitfield},
}

\newcommand{\qw}[1][-1]{\ar @{-} [0,#1]}
\newcommand{\qwx}[1][-1]{\ar @{-} [#1,0]}

\newcommand{\cw}[1][-1]{\ar @{=} [0,#1]}

\newcommand{\gate}[1]{*{\xy *+<.6em>{#1};p\save+LU;+RU **\dir{-}\restore\save+RU;+RD **\dir{-}\restore\save+RD;+LD **\dir{-}\restore\POS+LD;+LU **\dir{-}\endxy} \qw}
\newcommand{\meter}{\gate{\xy *!<0em,1.1em>h\cir<1.1em>{ur_dr},!U-<0em,.4em>;p+<.5em,.9em> **h\dir{-} \POS <-.6em,.4em> *{},<.6em,-.4em> *{} \endxy}}

\newcommand{\control}{*-=-{\bullet}}

\newcommand{\ctrl}[1]{\control \qwx[#1] \qw}
\newcommand{\rstick}[1]{*!L!<-.5em,0em>=<0em>{#1}}
\newcommand{\lstick}[1]{*!R!<.5em,0em>=<0em>{#1}}

\newcommand{\Qcircuit}{\xymatrix @*=<0em>}
\newcommand{\bra}[1]{\langle #1\vert}
\newcommand{\ket}[1]{\vert#1\rangle}

\newcommand{\id}{\mathbf{1}}

\newcommand{\eqref}[1]{Eq.~(\ref{#1})}
\begin{document}
\date{\today}
\title{\titlename}
\author{J. D. Whitfield} 
\email[email: ]{JDWhitfield@gmail.com}
\affiliation{Vienna Center for Quantum Science and Technology\\University of Vienna, Department of Physics, Boltzmanngasse 5, Vienna, Austria 1190}

\begin{abstract}
Time evolution of quantum systems is of interest in physics, in chemistry, and, more recently, in computer science. Quantum computers are suggested as one route to propagating quantum systems far more efficiently than ordinary numerical methods.  
In the past few years, researchers have actively been improving quantum simulation algorithms, especially those in second quantization.  
This work continues to advance the state-of-the-art by unifying several diverging approaches under a common framework.  In particular, it highlights the similarities and differences of the first and second quantized algorithms which are usually presented in a distinct fashion.  By combining aspects of the two approaches, this work moves towards an online second quantized algorithm operating within a single-Fock space.  This paper also unifies a host of approaches to algorithmic quantum measurement by removing superficial differences. The aim of the effort is not only to give a high-level understanding of quantum simulation, but to move towards experimentally realizable algorithms with practical applications in chemistry and beyond.
\end{abstract}

\maketitle

In the recent past, quantum information and quantum computing have had far reaching consequences for modern quantum chemistry. Quantum information, on the one hand, has inspired powerful new numerical methods. Entanglement based ansatzes collectively known as tensor network methods have become the primer method for obtaining ground states of one-dimensional and quasi-one-dimensional electronic system with promising applications to condensed matter \cite{Verstraete08} and chemical system \cite{Chan12,Marti10}. On other hand, new quantum algorithms contain much promise for efficient paths to simulating quantum dynamics as well as learning about ground state behavior. Quantum simulation is the idea of using quantum computational devices for more efficient simulation \cite{Feynman82}. Since the dynamics are simulated by a quantum system rather than calculated by a classical computer, quantum simulation often offers exponential advantage over classical simulation \cite{Georgescu14} for the calculation of electronic energies, reaction rates, correlation functions and molecular properties.

Although a series of algorithmic improvements in second quantized quantum simulation have come out~\cite{Mcclean14,Wecker14,*Babbush14,*Poulin15} focused on improving the algorithm presented in~\cite{Whitfield11}, none of these articles have attempted to leverage the group theoretic insights presented in Ref.~\cite{Whitfield13b}.  Here, the effort began in Ref.~\cite{Whitfield13b} is continued by attempting to construct a second quantized algorithm that acts only in a particular Fock space.  This has parallels with the locality arguments presented in Ref.~\cite{Mcclean14} but directly builds upon the machinery used for first quantized simulations \cite{Zalka98,Kassal08,Jones12}.  

To arrive at an algorithm that operates within a particular Fock space, various aspects of quantum simulations must be unified.  Consistent with the goal of unifying approaches to quantum simulation, various algorithms for quantum measurement are compared. Despite the various articles on this topic spanning decades~\cite{vonNeumann55,Ramsey63,Kitaev95,Cleve98,Zalka98,Biamonte11,Wang12}, there are only superficial differences between these approaches.  This is illustrated by showing equivalences among the various presentations of quantum measurement.  



The article begins with basic concepts from lattice quantum theory including the proper definitions of first and second quantization.  In the following section, two canonical quantum algorithms are compared and contrasted.  This naturally leads to a new approach to second quantized simulation. In the penultimate section, various schemes for quantum measurement are united into a single schema. A summary and outlook towards the future ends the article.

\section{Lattice quantum theory}\label{sec:latticeqm}
 Two ways to approach lattice systems are with first or second quantization.  The underlying lattice sites are completely arbitrary and could be, for example, Slater type orbitals, Gaussian type orbitals, quadrature points from classical polynomials, or plane waves.  For the ensuing discussion, the lattice selection is done by choosing the eigenbases of one-body operators. For the remainder of the article, $N$ is the number of particles and $M$ is the number of sites.
 
In first quantization, one deals with a $M^N$ dimensional complex tensor with $N$ indices denoted as $\Psi(x_1 x_2 ...x_N)\in \mathbf{C}$ with $x_i\in\{1,2\cdots,M\}$.  This must be a completely antisymmetric tensor meaning $\Psi(\cdots i\cdots j\cdots)=-\Psi(\cdots j \cdots i\cdots)$. Here capital letters are used to indicate collective indices i.e. $X=(x_1\cdots x_N)$. The wave function in bra-ket notion is:
\begin{equation}
	|{\Psi}\rangle=\sum_{x_1}^{M}\sum_{x_2}^M\cdots\sum_{x_N}^M\Psi(x_1x_2...x_N)\; |x_1...x_N\rangle=\sum_X^{M^N}\Psi(X)\ket{X}
\end{equation}
where vectors $|x_1x_2...x_N\rangle$ are the many-particle basis vectors of the $M^N$ dimension space. 

In second quantization, one has a listing of the unique elements of the first quantized tensor, that is a vector $D$ with $M\choose N$ complex elements.  The indexing of vector $D$ is given by ordered $N$-tuples.  One writes the wave function in second quantization as
\begin{equation}
	\ket{\Psi}=\sum_{K_1<K_2<...<K_N}^{ M\choose N} D_{K} \hat a_{K_1}^\dag \hat a_{K_2}^\dag ...|\Omega\rangle=\vec{\sum_K} \Psi_K \hat K ^\dag \ket{\Omega}
      \label{eq:2nd}
\end{equation}
The operators $\{\hat a_i,\hat a^\dag_j: 1\leq i,j \leq M\}$ satisfy $\hat a_i \hat a_j=-\hat a_j\hat a_i$ and $\hat a^\dag_j\hat a_i=\id_{ij}-\hat a_i\hat a^\dag_j$ where $\id_{ij}=\delta_{ij}$ whenever the underlying orbital basis is orthonormal. The vector $\ket\Omega$ is an arbitrary fixed vector in an $2^M$ dimensional space called the vacuum state and $\vec{\sum}$ indicates that the sum is over ordered tuples.

In lattice systems, first and second quantization are connected by local-basis transforms of the form $\hat b_x=\sum_j \hat a_j C_{jx}$. The anti-commutation relations between the two bases is given by $[\hat b_y, \hat a_i^\dag]_+=C_{iy}$.  To 
illustrate the connection, first define $ \hat X=\prod_{i=1}^N \hat b_i$ and $\hat X_m=\prod_{i\neq m}^N\hat b_i$ 
 \begin{eqnarray}
\langle \hat X \hat J^\dag
\rangle&=&C_{j_1b_1}
\langle\hat X_1\hat J_1^\dag
\rangle
-\langle  ... \hat b_2\hat a_{j_1}^\dag\hat b_1  \hat a_{j_2}^\dag...
\rangle \\
&=& C_{j_1b_1}
\langle\hat X_1\hat J_1^\dag
\rangle
-C_{j_1b_2}\langle\hat X_2\hat J^\dag_1\rangle\nonumber\\
&&+\langle ... \hat b_3 \hat a_{j_1}^\dag \hat b_2\hat b_1\hat a_{j_2}^\dag...\rangle\\
&=&\sum_m^N (-1)^{m+1} C_{j_1b_m}
\langle\hat X_m\hat J_1^\dag
\rangle\\
&=&\left|
\begin{array}{ccc} 
C_{j_1b_1} &\cdots & C_{j_1 b_N}\\
\vdots &\ddots &\vdots\\
C_{j_Nb_1}&\cdots&C_{j_Nb_N}
\end{array}
\right|\label{eq:det}
\end{eqnarray}
The final expression is called the Pl\"ucker embedding of matrix $C$ into an antisymmetric space \cite{Schilling14}.  Writing $C_{jx}$ as $\phi_j(x)$, one sees that this is the standard Slater determinants from quantum chemistry up to the normalization factor $1/ \sqrt{N!}$ which accounts for local basis functions independently normalized.  

\subsubsection*{Hamiltonians}
In non-relativistic electronic quantum mechanics, the Hamiltonian consists of kinetic energy, two-electron Coulomb interactions, and a scalar field representing the electronic environment e.g.~the nuclear charges. 

In a lattice basis corresponding to eigenfunctions of the potential operator,
the potential energy operator is $\hat V=\sum V_r\hat  a_r^\dag\hat  a_r$ and the kinetic energy operator is $\hat T=\sum_{pq}^M T_{pq} \hat a_p^\dagger\hat a_q$. Here, it is needed that the $M$ x $M$ matrix $T$ is a fixed but arbitrary real symmetric matrix. The kinetic energy operator is complex if time reversal symmetry is broken. Define momentum space as the basis where $T$ is diagonal and let   $\{\hat b_p\}$ be the corresponding anti-symmetric operators.  

In the continuum limit, the change of basis between $\hat a_i$ and $\hat b_j$ is via the Fourier transform. However, in a discrete basis, this may not be the case due to various methods for approximating the kinetic energy operator. Still, the position and momentum spaces are related by some local unitary transform. 

An electron-electron interaction is written as $\hat W=\frac12 \sum W_{pqrs}\hat a_p^\dag\hat a_q^\dag\hat a_r\hat a_s$. For the Coulomb interaction, this operator is diagonal in the position basis i.e. $\hat W=\frac12 \sum W_{pq}\hat a_p^\dag\hat a_q^\dag\hat a_q\hat a_p$.  Moreover, the integral $W_{pq}$ depends only on the distance between sites $p$ and $q$. Note that all of the Hamiltonian terms are permutationally invariant.

\section{Unified Quantum simulations}
In this section, we will unify the first and second quantized algorithms in a common framework and then create an partially online second quantized algorithm. By online, it is meant that the computation is done coherently in the quantum hardware. Before turning to the main thrust of the paper, it is worth mentioning algorithms simulating spare Hamiltonians following Refs. \cite{Berry07,Aharonov08} have recently been exploited in the context of quantum chemical simulations by Ref.~\cite{Toloui13}.  The ideas presented below may also be applicable in this setting.

In first quantized quantum simulations\cite{Zalka98,Kassal08,Jones12}, the system is evolved under the operators $\hat W+\hat V$ and $\hat T$ in the position and momentum basis, respectively. The quantum Fourier transform is used to efficiently convert between position space and momentum space. The algorithms requires $N\log M$ qubits to store the wave function and evolves the system only in a single Fock space. Antisymmetry is only imposed on the initial state which is stored in a binary representation whereby $m$ qubits can store $M=2^m$ sites. In this work, this is the \textbf{A1} algorithm.

In second quantized quantum simulations, the system is evolved under the Hamiltonian $H=\sum (T_{pq}+V_{pq})a_p^\dag a_q+ \frac12\sum W_{pqrs}\hat a_p^\dag\hat a_q^\dag\hat a_r\hat a_s$.  In the standard algorithms \cite{Whitfield11,Wecker14,*Babbush14,*Poulin15}, the integrals $\{V_{pq},T_{pq},W_{pqrs}\}$ are precomputed classically and the wave function is stored in unary representation where $M$ qubits are needed for $M$ sites. The exponentially large Fock space (with a total of $\sum_k {M\choose k} =2^M$ states) is only used for simulations that require changes in particle number such as super-conducting interactions approximated by $a_ia_j + a_i^\dag a_j^\dag$, grand canonical simulations with Hamiltonian terms $\mu \sum_k a_k^\dag a_k$, or electron affinity/ionization calculations. When the particle number is fixed to $N$ electrons, the evolution occurs in a subspace with only ${M \choose N}$ states.  \textbf{A2} is used to refer to this algorithm.

\subsubsection{Online second quantized simulation algorithm}
The desirability of simulating within a single Fock space was highlighted in Ref.~\cite{Whitfield13b}. The first steps towards a second quantized algorithm achieving that ultimate goal is presented. The insights stems from contrast that scaling of algorithm \textbf{A2} does not depend on the number of electrons being simulated whereas \textbf{A1} does.   The goal of simulating in a single Fock will not be fully achieved here, but similar to Ref.~\cite{Mcclean14}, a scaling of $O(M^2)$ is also obtained, but for very different reasons.  Ref.~\cite{Mcclean14} attempts to exploit locality of the local basis in order to show that the number of non-trivial Coulomb integrals scale as $O(M^2)$;whereas, here the techniques from \textbf{A1} are not completely adapted to second quantization.  For this reason, the kinetic energy operator needs to be implemented in all Fock spaces giving $O(M^2)$ contributions.  The spatial terms will scale with the number of electrons rather than the number of basis functions.

In \textbf{A1}, the algorithm computes all quantities on-the-fly using phase-kickback of classically computable quantities. The key ideas is to work in the shared eigenbasis of the scalar potential and the two-body interaction term.  Computing the Hamiltonian terms must be done coherently using quantum version of classical algorithms such as addition or multiplication \cite{Jones12, Kassal08}.  At a high-level, the two-electron term is computed as e.g.~follows:
\begin{eqnarray}
	&&\ket{scratch}\ket{110}\ket{011}\\
	&\mapsto&\left|{\frac1{d(3,6)}}\right\rangle\ket{110}\ket{011}\label{eq:1}\\
	&\mapsto&e^{-i\xi/d(3,6)}\left|{\frac1{d(3,6)}}\right\rangle\ket{110}\ket{011}\label{eq:2}
\end{eqnarray}
In \eqref{eq:1}, two points of clarification are useful: (1) the position is stored in binary representation and (2) a quantum version of a classical computation is effected.  For the implementation of the Coulomb interaction potential, the algorithm is predicated on addition, multiplication, and a Newton-Raphson method for implementing square roots leading to {$O(b^2N^2)$} scaling with $b$ binary bits of precision used for computation~\cite{Kassal08}.  In \eqref{eq:2}, the phase kickback method is used~\cite{Cleve98,Jones12}. 

The interesting point uncovered here is that this computation can be easily adapted for second quantized algorithms.  Indeed, since the Hamiltonian is permutationally invariant, the previous method can be directly adapted as,
\begin{eqnarray}
	&&\ket{scratch}\ket{00100100}\\
	&\mapsto&\left|{\frac1{d(3,6)}}\right\rangle\ket{00100100}\\
	&\mapsto&e^{-i\xi/d(3,6)}\left|{\frac1{d(3,6)}}\right\rangle\ket{00100100}
\end{eqnarray}
Note that here the positions are stored in unary representation where the third bit corresponds to the third position and that counting pairs of fermions is not difficult even when the particle number increases beyond two.

One of the key differences between \textbf{A1} and \textbf{A2}, is the exploitation of Fourier transformation to efficiently implement the kinetic energy operator in its eigenbasis.  In \textbf{A1}, this change of basis is done via the quantum Fourier transform.  In \textbf{A2}, there are two obstacles: First, the quantum Fourier transform is conducted in the binary representation rather than the unary representation used in \textbf{A2}.  Second, the coefficients of the Fourier transform must be anti-symmetrized as in \eqref{eq:det}.  These difficulties are probably amendable using conversions from unary to binary, quantum Fourier transform, and the reversible anti-symmetrization algorithm~\cite{Abrams97}.  However, this naive approach seems to defeat many of the key advantages of the \textbf{A2}. Accordingly, an elegant solution to performing local change of basis in second quantization is highly desirable in order to exploit the simplicity of the kinetic energy in the momentum basis.

Despite the inability to efficiently convert between local bases in second quantized simulations, this paper prescribes an approach where by only the kinetic energy is implemented as in previous works \cite{Whitfield11,Wecker14,*Babbush14,*Poulin15} but the potential and the two-body part of the Hamiltonian are computed using their shared eigenbasis with the algorithm adapted from \textbf{A1} as presented here. This online version of the second quantized algorithm requires only the pre-computation of the kinetic energy operator. The online portions in the position basis can may be sped up by quantum pre-computation \cite{Jones12}.  The detailed cost analysis of the computations required by the online version was previously done in \cite{Jones12}.  The full comparison of the merits of the standard second quantized algorithm and the online version introduced here will appear in the near future. 

\section{Unified quantum measurements}
Thus far, the focus has been on the algorithms for propagating the quantum system.  Now, attention turns to the task of measuring properties of the system.  Several schemes are considered and shown to be equivalent.  In order of appearance, the methods are von Neumann's measurement apparatus \cite{vonNeumann55,Zalka98}, Ramsey spectroscopy \cite{Ramsey63},  phase estimation algorithm \cite{Kitaev95}, phase kickback \cite{Cleve98}, adiabatic quantum simulation \cite{Biamonte11}, and the energy spectrum algorithm \cite{Wang12}.  All of these schemes are the same up to superficial differences.  

The earliest example of the quantum measurement scheme was attributed to von Neumann by Ref.~ \cite{Zalka98}. There one has an interaction Hamiltonian $H_{\text{int}}=\hat A\otimes \hat p$ where $\hat A=\sum\lambda_a\ket{a}\bra{a}$ is the observable to be measured.  With initial state $\ket{\alpha}\otimes\ket{x=0}=\ket{\alpha}\otimes a_0^\dag\ket{\Omega}$, the evolution for a time $2\pi/M$ yields 
\begin{equation}
	e^{-2\pi i \hat A\hat p/M}\ket{\alpha}\ket{0}=\ket{\alpha}e^{-2\pi i\lambda_\alpha/M \hat p}\ket{x=0}=\ket{\alpha}\ket{x=\lambda_\alpha}
\end{equation}
To obtain this result, the definitions from the Section \ref{sec:latticeqm} are used with $\hat p=\sum k \hat b_k^\dag \hat b_k$ and change of basis between position and momentum basis defined by $C_{kn}=\exp(-2\pi i kn/M)/\sqrt{M}$. 

In the quantum information community, the most well known example of quantum measurement is the one given by \cite{Kitaev95}:
\begin{equation}
\centerline{\Qcircuit @C=1em @R=1em @!R {
	\lstick{\ket{0}}	              	&\gate{\sf H}	&\ctrl{1}      	 &\gate{\sf H}	& \meter 	 &\rstick{b_n} \cw\\
	\lstick{\ket{\alpha}}	&\qw      	&\gate{U}	&\qw      	&\qw		&\rstick{\ket{\alpha}} \qw
        }}
\end{equation}
where $\sf H$ is the Hadamard transform defined as the Fourier transform over $M=2$ sites.

Next, analysis shows that Kitaev's scheme and the von Neumann measurements are essentially the same. Kitaev's scheme requires implementing a controlled $U$.  In equations, this is $\ket{0}\bra{0}\otimes \mathbf{1} +\ket{1}\bra{1}\otimes U$.   Since this matrix is block diagonal, its generator is as well.  The generator of the identity is just the zero matrix and the generator of $U$ is the Hamiltonian whose spectrum is of interest, $\hat A$.  The generator of the controlled unitary is then $H_{\text{int}}=\ket{1}\bra{1}\otimes \hat A$.  This form is the same as is found in schemes of \cite{Biamonte11} and \cite{Wang12}.

To fully illustrate the point, let us focus on the von Neumann scheme described above.  Consider the case that there are only two sites i.e.~$M=2$.  The kinetic energy operator for a two site system is proportional to the Hadamard (Fourier) transform of $\sigma^z$
\begin{equation}
\sigma^x=\left( \begin{array}{cc} 0 &1\\1&0\end{array}\right)=\frac1{\sqrt{2}}\left( \begin{array}{cr} 1 &1\\1&-1\end{array}\right) \left( \begin{array}{cc} 1 &\\&-1\end{array}\right) \frac1{\sqrt{2}}\left( \begin{array}{cr} 1 &1\\1&-1\end{array}\right)
\end{equation}
Shifting and rescaling the kinetic energy such that $\hat p=\frac{1}{2}(\mathbf{1}-\sigma^x)$, one can rewrite the coupling Hamiltonian as 
\begin{equation}
\tilde H=\hat p\otimes \hat A=({\sf H}\otimes \mathbf{1})  \left(\frac{\mathbf{1}-\sigma^z}{2}\otimes \hat A\right) ({\sf H}\otimes \mathbf{1}).
\end{equation}
This Hamiltonian is precisely the generator of the Kitaev circuit.

It remains to show that Ramsey spectroscopy \cite{Ramsey63} is equivalent to the other schemes.  Ramsey spectroscopy is used to measure an unknown energy splitting.  The protocol begins with a $\pi/2$ pulse, followed by free evolution for time $T$, and ends with a final $\pi/2$ pulse.  The population is transfered to a phase difference and after some time evolution, the change in phase due to energy differences can be converted back into population.  

To show equivalence, the idea of ``phase kickback,''  which is yet another description of the von Neumann scheme, introduced by Ref.~\cite{Cleve98} is utilized.
Because the state $\ket{\alpha}$ is an eigenvector of $\hat A$, the evolution under $\hat A$ does nothing but impart a phase on the controlled register.  In our case, this is what sets the qubit frequency $\omega_0$.  This phase is said to be ``kicked'' back onto the control qubit (hence the name).  From a reductionist point of view, the wave function register is superfluous and one can write the Kitaev circuit with a phase gate on the probe register:
\begin{equation}
\centerline{
	\Qcircuit @C=1em @R=1em @!R {
	\lstick{\ket{0}}	              	&\gate{\sf H}	&\gate{\sf G(\lambda_\alpha t)}      	 &\gate{\sf H}	& \meter 	 &\rstick{b_n} \cw
        }
}
\end{equation}	
Here $G(\theta)=\ket{0}\bra{0}+\exp(-i\theta )\ket{1}\bra{1}$.  To see the equivalence with the other schemes, consider the generator of such a phase gate $H_{G}$. For $H_G$, one needs $\bra{0}H_G\ket{0}=0$ and $\bra{1}H_G\ket{1}=\lambda_\alpha$.  This is nearly identical to the Ramsey spectroscopic scheme.  The minor difference is accounted for by the fact that the $\pi/2$ pulse is not exactly the same as the Hadamard transformation. This is merely selecting a different approximation to the kinetic energy operator, but the outcomes are the same with the probability of being the $\ket{1}$ state being $\frac12 (1+\cos (\lambda_\alpha t))$ in both schemes.

It should be noted that the exact number of measurements needed to estimate the value of a parameter \cite{Griffiths96,Aspuru05,Svore13,Chiang14} depends on how past measurements are utilized for the current measurement, what choice is made for the momentum operator, and how the inference is done. Thus, while the form of the measurements does not change, the usage of information obtained from measurements can be optimized. 

Further, note that the eigenvalues are always measured relative to something.  It is never an absolute energy and this is a simple consequence of Fourier theory.  Stated differently, the Hamiltonian operators can be unbounded but the unitary group is compact.  This compactness means that you cannot see the full spectrum and that eigenvalues must eventually "wrap around."   The aliasing of the eigenvalues arises because the dynamics are unaffected by a shift of energy. Hence, there is a gauge degree for freedom which can be arbitrarily selected e.g.~as all charges infinitely separated in electrostatics or all atoms completely separated in thermo-chemistry. A through discussion of aliasing in the context of quantum phase estimation is can be found in Ref.~\cite{Whitfield11}.

From Hamiltonian complexity results \cite{Whitfield13a, Osborne12}, the quantum measurement of an eigenvalue is not an easy task.  This is because Hamiltonians are exponentially larger than the problem size, hence there are an exponential number of eigenstates.  If one wishes to scan the ``absorption'' spectrum by changing the probe's energy as in the scheme of Ref. \cite{Wang12}, the input states must have polynomially large overlap with the eigenstates for peaks to occur at eigenvalues. Otherwise, it will take an exponential number of measurements to see enough tunneling events to determine the transition energy.

\section{Outlook}
This article focused on understanding and combining various aspects of existing quantum simulation algorithms.  It was also shown that, despite some information-based improvements, the von Neumann scheme for quantum measurement has been re-discovered many times over the half century since its introduction.  The introduction of a partially online second quantized algorithm opens the door to a fully online second quantized quantum simulation but this will require an intelligent method for changing the local basis from position space (eigenbasis of $\hat V$) to momentum space (eigenbasis of $\hat T$). 

Many experimental groups worldwide are pushing for more precise control over larger realizations of quantum computers, but simulations that enable quantum computation to produce results unobtainable by standard numerical methods have yet to appear.  As a model chemistry \cite{Pople99}, quantum simulation is still stuck in the third of the five stages of development: (1) target accuracy, (2) formulation, (3) implementation, (4) validation, and (5) prediction. 

The current work continues to simplify the formulation in an effort to make the implementation more feasible for experimental technologies. But to move on to the fourth stage, a test set of molecular instances must be tested in functional quantum hardware. Until then, the ultimate goal of making predictions will remain far into the future.

Taking an eye towards near-future schemes for quantum simulation, 
a paradigm shift from competition to collaboration is needed. Instead of asking how a quantum
computer can beat a classical computer, we should be asking how a quantum sub-processor can
enhance commonplace processors. The necessary developments to bring this to fruition are (1) applications where small quantum processors can outperform small CPUs, (2) applications that require small, short quantum evolutions as part of a larger application (e.g. crucial quantum coherent subsystems) and (3) a quantum-classical data bus. These are the components that will make quantum computation a viable technology that will become part of commercial products. This shifts the emphasis from larger quantum computers to faster, more reliable ones and that should only require modest advances in experimental technology.

\begin{acknowledgments}
\paragraph{Acknowledgments}
I would like to thank the VCQ Fellowship for funding and support.  Discussions with R.~Babbush inspired the comments on quantum measurement algorithms. 
\end{acknowledgments}

\end{document}